\documentclass[showpacs,preprintnumbers,amsmath,aps,nofootinbib,amssymb,superscriptaddress]{revtex4}
\usepackage{graphicx}
\usepackage{dcolumn}
\usepackage[english]{babel}
\usepackage{bm}
\usepackage[font=scriptsize]{caption}
\usepackage{ragged2e}
\usepackage[font=small,labelfont=bf,justification=raggedright]{caption}
\usepackage[utf8]{inputenc}
\usepackage[tikz]{bclogo}
\usepackage[framemethod=tikz]{mdframed}
\usepackage{amsmath}

\usepackage{mathtools}
\usepackage{xcolor}
\usepackage{lscape}
\usepackage{rotating}
\usepackage{hyperref}
\colorlet{BLUE}{blue} \colorlet{RED}{red}
\newsavebox{\measurebox}

\begin{document}

\title[]{Quintom fields from chiral  anisotropic cosmology}
\author{J. Socorro}
\email{socorro@fisica.ugto.mx}
 \affiliation{Departamento de
F\'{\i}sica, DCeI, Universidad de Guanajuato-Campus Le\'on, C.P.
37150, Le\'on, Guanajuato, M\'exico}

\author{S. P\'erez-Pay\'an}
\email{saperezp@ipn.mx} \affiliation{Unidad Profesional
Interdisciplinaria de Ingenier\'ia,
Campus Guana\-jua\-to del Instituto Polit\'ecnico Nacional.\\
Av. Mineral de Valenciana \#200, Col. Fraccionamiento Industrial
Puerto Interior, C.P. 36275, Silao de la Victoria, Guana\-juato,
M\'exico.}

\author{Rafael Hern\'andez-Jim\'enez}
\email{rafaelhernandezjmz@gmail.com}
 \affiliation{Departamento de
F\'isica, Centro Universitario de Ciencias
Exactas e Ingenier\'ia, Universidad de Guadalajara.\\
Av. Revoluci\'on 1500, Colonia Ol\'impica C.P. 44430, Guadalajara,
Jalisco, M\'exico.}

\author{Abraham Espinoza-Garc\'ia}
\email{aespinoza@ipn.mx} \affiliation{Unidad Profesional
Interdisciplinaria de Ingenier\'ia,
Campus Guana\-jua\-to del Instituto Polit\'ecnico Nacional.\\
Av. Mineral de Valenciana \#200, Col. Fraccionamiento Industrial
Puerto Interior, C.P. 36275, Silao de la Victoria, Guana\-juato,
M\'exico.}

\author{Luis Rey D\'iaz-Barr\'on}
\email{lrdiaz@ipn.mx} \affiliation{Unidad Profesional
Interdisciplinaria de Ingenier\'ia,
Campus Guana\-jua\-to del Instituto Polit\'ecnico Nacional.\\
Av. Mineral de Valenciana \#200, Col. Fraccionamiento Industrial
Puerto Interior, C.P. 36275, Silao de la Victoria, Guana\-juato,
M\'exico.}

\begin{abstract}

In this paper we present an analysis of a chiral anisotropic cosmological scenario from the perspective of quintom fields. In this setup quintessence and phantom fields interact in a non-standard (chiral) way within an anisotropic Bianchi type I background. We present our examination from two fronts: classical and quantum approaches. In the classical program we find analytical solutions given by a particular choice of the emerged relevant parameters. Remarkably, we present an explanation of the ``big-bang'' singularity by means of a ``big-bounce''. Moreover, isotropization is in fact reached as the time evolves. On the quantum counterpart the Wheeler-DeWitt equation is analytically solved for various instances given by the same parameter space from the classical study, and we also include the factor ordering $\rm Q$. Having solutions in this scheme we compute the probability density, which is in effect damped as the volume function and the scalar fields evolve; and it also tends towards a flat FLRW framework when the factor ordering constant $\rm Q\ll 0$. This result might indicate that for a fixed set of parameters, the anisotropies quantum-mechanically vanish for very small values of the parameter $\rm Q$. Finally, classical and quantum solutions reduce to their flat FLRW counterparts when the anisotropies vanish.

\end{abstract}

\maketitle
\section{Introduction} 
The rather small deviation from isotropy observed in the cosmic microwave background (CMB) radiation \cite{Martinez:1995} makes it plausible that at very early times the universe was indeed anisotropic, therefore prompting the introduction of anisotropic cosmological models to describe the evolution of the universe near the initial singularity~\cite{Belinskii:1972, Folomeev:2000}. The Bianchi type I model is a natural choice for such a background given that its isotropic limit is the spatially flat Friedmann-Robertson-Lama$\hat{\i}$tre-Walker (FRLW) model (see, e.g., \cite{Ryan}). Indeed, the Bianchi type I model has been recently considered to explain the aforementioned tiny variations in the CMB by a number of researchers \cite{Amirhashchi:2018, Amirhashchi:2019, Akarsu et al:2019, Goswami et al:2020}.

On the other hand, the multi-field cosmology paradigm has proven to be an effective framework to account (in a single model) for several important characteristics/ingredients of the universe, e.g., early acceleration (inflation)~\cite{Sasaki:1995aw, Liddle:1998jc, Rigopoulos:2002mc, Bassett:2005xm, Wands:2007bd, Lalak:2007vi, Ashoorioon:2009wa, Achucarro:2010jv, Achucarro:2010da, Achucarro:2012sm, Achucarro:2012yr, Pi:2012gf, Renaux-Petel:2015mga, Brown:2017osf, Achucarro:2017ing, Achucarro:2018vey, Aragam:2020uqi}, dark matter~\cite{Hu:2000ke, Hui:2016ltb, Tellez-Tovar:2021mge}, late acceleration~\cite{Boyle:2001du, Kim:2005ne, vandeBruck:2009gp, BeltranJimenez:2012xud, Vardanyan:2015oha, Leithes:2016xyh, Akrami:2017cir, Cicoli:2020cfj, Cicoli:2020noz, Akrami:2020zfz, Socorro:2020nsm, Paliathanasis:2021fxi, Burgess:2021qti, Burgess:2021obw, Anguelova:2021jxu, Elizalde:2004mq, Motoa-Manzano:2020mwe, Orjuela-Quintana:2020klr}. 
With respect to early and late acceleration, the crossing of the phantom divide line is a most wanted feature in scalar field cosmology; it has been shown that this crossing cannot be achieved by considering a single scalar field/fluid (unless stability is not demanded) \cite{Cai:2009zp}. The standard quintom scenario \cite{Feng:2005} considers two scalar fields, a quintessence and a phantom, in order to realize such crossing in a simple way. As a byproduct, quintom fields allow (in particular cases) the avoidance of the initial singularity by means of a bounce~\cite{Socorro:2022aoz} (see also the review \cite{Cai:2009zp}). In the conventional quintom scenario (and in ordinary multi-field cosmology)
the scalar fields interact in the following way:
\begin{equation}
\mathcal{L}_\phi=\delta^{ab}g^{\mu\nu}\nabla_\mu\phi_a\nabla_\nu\phi_b+V(\phi_a,\phi_b) \,.\label{stnd-multifield}
\end{equation}
An incarnation of multi-field cosmology is the so called chiral-cosmology \cite{chervon1995}, in which the scalar fields define an ``internal space'' with a certain metric component $m_{ab}$. They also interact in a non-standard manner within their kinetic terms, their couplings are governed by the metric $m_{ab}$ (in short, we will replace $\delta^{ab}\to m^{ab}$ in \eqref{stnd-multifield}). This metric can be seen as arising from casting a non-minimally coupled multi-scalar-tensor theory as General Relativity (i.e., in going from the Jordan frame to the Einstein frame) \cite{Kaiser:2010}. Non-minimal couplings are indeed required when considering the quantization of scalar fields in curved backgrounds \cite{Birrell}, the use of non-canonical fields in (effective) descriptions of the early universe in Einstein's general relativity is therefore theoretically consistent with standard quantum field theory.

In the present investigation we consider a Bianchi type I framework within a generalized quintom scenario, in which the scalar fields define a chiral space with a certain metric $m_{ab}$ (so that the fields are not canonical). The exact classical solutions are obtained, then with them we show that the initial singularity is avoided by means of a bounce. Moreover, exact quantum solutions will show that the wave function of the universe is damped with respect to the average scale factor. Similar conclusions were made in the corresponding isotropic case~\cite{s-2021}. In the remaining part of this introduction we proceed to describe the generalities of the
chiral cosmology which we will be employing. We consider the following simple case of two scalar fields, a {\bf quin}tessence field $\phi_1$ and phan{\bf tom} field $\phi_2$ (with their corresponding potential terms) within the chiral cosmology paradigm
\cite{s-2021,chervon1995,Chervon2013,Chervon2015,Fomin2017,Paliathanasis2019,chervon:2019,Fomin:2021}
\begin{equation}
\rm {\cal L}=\sqrt{-g} \left( R + m^{ab}\,  \xi_{ab}(\phi_c,g^{\mu
    \nu})+ C(\phi_c)\right) \,, \label{lagra}
\end{equation}
where $\rm R$ is the Ricci scalar, $\rm
\xi_{ab}(\phi_c,g^{\mu \nu})=-\frac{1}{2}g^{\mu \nu}\nabla_\mu
\phi_a \nabla_\nu \phi_b$ the kinetic energy, and $\rm C(\phi_c)= V(\phi_a,\phi_b)$ the corresponding scalar field potential, with $\rm m^{ab}$ a $2 \times 2$ constant matrix; we consider the particular form
$\rm m^{ab}=\left(
\begin{tabular}{cc}
$\rm 1$ & $\rm m^{12}$ \\
$\rm m^{12}$ & $\rm -1$ \end{tabular}
\right).$
Thus, the Einstein-Klein-Gordon field equations are
\begin{equation}
\rm G_{\alpha \beta}=\rm -\frac{1}{2}m^{ab} \left(\nabla_\alpha
\phi_a \nabla_\beta \phi_b -\frac{1}{2}g_{\alpha \beta} g^{\mu \nu}
\nabla_\mu \phi_a \nabla_\nu \phi_b \right) +\frac{1}{2}g_{\alpha
    \beta} \, C(\phi_c) \,, \label{mono}
\end{equation}
\begin{equation}\rm
m^{cb} \nabla_\nu \nabla^\nu \phi_b -\frac{\partial
    C(\phi_c)}{\partial \phi_c}=0 \,, \label{K-G}
\end{equation}
where $a,b,c=1,2$. From (\ref{mono}) we read off the energy-momentum
tensor for the scalar fields $\rm (\phi_1,\phi_2)$, as
\begin{equation}
\rm 8\pi G T_{\alpha \beta}(\phi_1,\phi_2)=-\frac{1}{2}m^{ab}
\left(\nabla_\alpha \phi_a \nabla_\beta \phi_b -\frac{1}{2}g_{\alpha
    \beta} g^{\mu \nu} \nabla_\mu \phi_a \nabla_\nu \phi_b \right)
+\frac{1}{2}g_{\alpha \beta} \, V(\phi_1,\phi_2) \,. \label{tmunu}
\end{equation}
and considering the analogy with a barotropic perfect
fluid for the scalar fields,
\begin{equation}
\rm T_{\alpha \beta}(\phi_c)=(\rho + P)u_\alpha(\phi_c)
u_\beta(\phi_c) + P\, g_{\alpha \beta} \,,
\end{equation}
we have that the pressure P and the energy density $\rm\rho$ of the
scalar fields are
\begin{equation} 
\rm
P(\phi_c)=\frac{1}{2}m^{ab}\,  \xi_{ab} -\frac{1}{2}C(\phi_c),
\qquad \rho(\phi_c)=\frac{1}{2}m^{ab}\,  \xi_{ab} +
\frac{1}{2}C(\phi_c) \,,
\end{equation}
the four-velocity becomes $\rm u_\alpha u_\beta=\frac{\nabla_\alpha\phi_a \nabla_\beta \phi_b}{2\xi_{ab}}$. 

We will employ the scalar potential term $\rm
C(\phi_c)=V_1(\phi_1)+V_2(\phi_2)=V_{01}e^{-\lambda_1\phi_1}+V_{02}e^{-\lambda_2 \phi_2}$ (with $\lambda_1$, $\lambda_2$ non-negative) and the line element to be considered for this two-field cosmological model will be that of the anisotropic Bianchi type I model, which in Misner's parameterization is given by
\begin{equation}
\rm ds^2=-N^2 dt^2 +e^{2\Omega + 2\beta_+ + 2\sqrt{3}\beta_-}dx^2
+e^{2\Omega + 2\beta_+ - 2\sqrt{3}\beta_-}dy^2+e^{2\Omega -
    4\beta_+}dz^2 \,, \label{metric}
\end{equation}
where the scale factors are $\rm A=e^{\Omega+\beta_+ +\sqrt{3}\beta_-}, B=e^{\Omega + \beta_+ -\sqrt{3}\beta_-}, C=e^{\Omega - 2\beta_+}$ and $(\beta_+,\beta_-)$ are the anisotropic parameters. Also $(\Omega,\beta_+,\beta_-)$ are scalar
functions depending on time, and $\rm N=N(t)$ is the lapse function. Plugging in (\ref{metric}) into (\ref{lagra}) we obtain the following Lagrangian density (we eliminate the second time derivatives, previously)
\begin{equation}
\rm {\cal L}=e^{3\Omega}\left\{6\frac{\left( \dot \Omega
    \right)^2}{N} -6\frac{\left( \dot \beta_+
    \right)^2}{N}-6\frac{\left( \dot \beta_- \right)^2}{N} -
\frac{(\dot \phi_2)^2}{2N}-\frac{m^{12}}{N}\dot \phi_1 \dot
\phi_2+\frac{(\dot \phi_2)^2}{2N}+N\left(
V_1(\phi_1)+V_2(\phi_2)\right)\right\}. \label{lagrangian}
\end{equation}
Henceforth we will be utilizing the Lagrangian density~(\ref{lagrangian}) as starting point for our study. The document is organized as follows. Section II is devoted to set up the classical scheme via the Hamiltonian formalism, and to obtain exact classical solutions for several cases. In section III the Wheeler-DeWiit equation is constructed considering a semi-general factor ordering, and exact quantum solutions are presented for various cases as well. Final remarks are stated in section IV.

\section{Classical scheme}
In this section we present the classical solutions via the Hamiltonian formalism. We start with the momenta $\rm \Pi_q=\partial {\cal
L}/\partial\dot q$ (with $\rm
q^i=\Omega,\beta_+,\beta_-,\phi_1,\phi_2$), which are calculated in the usual way, yielding
\begin{equation}\label{momenta}
\begin{split}
\rm \Pi_\Omega &=\rm \frac{12}{N}e^{3\Omega}\dot \Omega,\\
\rm \Pi_+ &=\rm - \frac{12}{N}e^{3\Omega}\dot \beta_+ ,\\
\rm \Pi_- &=\rm - \frac{12}{N}e^{3\Omega}\dot \beta_- ,\\
\rm\Pi_{\phi_1} &=\rm \frac{1}{N}e^{3\Omega}(-\dot \phi_1- m^{12}\dot\phi_2),\qquad\\
\rm \Pi_{\phi_2} &=\rm \frac{1}{N}e^{3\Omega}(\dot \phi_2-
m^{12}\dot \phi_1),\\ 
\end{split}
\begin{split}
\rm\dot\Omega&=\rm\frac{N}{12}e^{-3\Omega}\Pi_\Omega,\\
\dot\beta_+&=\rm -\frac{N}{12}e^{-3\Omega} \Pi_+ ,\\
\dot \beta_-&=\rm -\frac{N}{12}e^{-3\Omega} \Pi_- ,\\
\dot\phi_1&=\rm -e^{-3\Omega}\frac{N}{1+(m^{12})^2}\left(\Pi_{\phi_1}+m^{12}\Pi_{\phi_2}\right),\\
\dot\phi_2&=\rm e^{-3\Omega}\frac{N}{1+(m^{12})^2}\left(m^{12}\Pi_{\phi_1}+\Pi_{\phi_2}\right),
\end{split}
\end{equation}
then the Lagrangian density (\ref{lagrangian}) is rewritten in a canonical way, i.e. $\rm {\cal L}_{canonical}=\Pi_i \dot q^i - N {\cal H}$, so we arrive at the Hamiltonian density
\begin{equation}
\rm {\cal H}=\frac{e^{-3\Omega}}{24}\left\{
\Pi_\Omega^2-\Pi_+^2-\Pi_-^2-12\frac{\Pi_{\phi_1}^2}{\triangle}-12\,m^{12}\frac{\Pi_{\phi_1}\Pi_{\phi_2}}{\triangle}
+12\frac{\Pi_{\phi_2}^2}{\triangle}-24V_{01}\,e^{6\Omega-\lambda_1
\phi_1} -24V_{02}\,e^{6\Omega-\lambda_2
\phi_2}\right\} \,,\label{hamiltonian}
\end{equation}
where $\triangle=1+(m^{12})^2$. We now consider the canonical transformation
$\rm(\Omega,\phi_1,\phi_2,\beta_+,\beta_-)\leftrightarrow
(\xi_1,\xi_2,\xi_3,\xi_4,\xi_5)$
\begin{equation}\label{trans_2}
\begin{split}
\rm \xi_1&=6\Omega-\lambda_1 \phi_1 \,,\\
\xi_2&= 6 \Omega-\lambda_2 \phi_2 \,,\\
 \xi_3&=6 \Omega + \lambda_1
\phi_1 + \lambda_2\phi_2 \,,\\
\xi_4&=\beta_+ \,,\\
\xi_5&=\beta_- \,,\\
\end{split}
\quad\longleftrightarrow\quad
\begin{split}
\Omega&=\rm\frac{\xi_1 + \xi_2+  \xi_3}{18} \,,\\
 \rm \phi_1&= \rm \frac{-2\xi_1 + \xi_2+\xi_3}{3\lambda_1} \,,\\
\rm \phi_2 &= \rm \frac{\xi_1-2\xi_2+\xi_3}{3\lambda_2} \,,\\ 
\rm \beta_+&=\xi_4 \,,\\
\rm \beta_-&=\xi_5 \,,
\end{split}
\end{equation}
with the new conjugate momenta $\rm (P_1,P_2,P_3,P_4,P_5)$ given by
\begin{eqnarray}\label{new-moment}
\rm \Pi_\Omega &=& \rm 6 P_1 +6 P_2+ 6 P_3 \,, \nonumber\\
\rm \Pi_{\phi_1} &=& \rm \lambda_1 \left(-P_1 + P_3\right) \,,\nonumber\\
\rm \Pi_{\phi_2} &=& \rm \lambda_2\left(- P_2 + P_3\right) \,,\\
\rm \Pi_+&=& \rm P_4 \,,\nonumber\\
\rm \Pi_-&=& \rm P_5 \nonumber \,. 
\end{eqnarray}
Therefore the Hamiltonian density, in the gauge $\rm N=24e^{3\Omega}$, becomes
\begin{multline}\label{hami-bi}
\rm {\cal H} = \rm  12 \left(3-\Lambda_1\right)P_1^2+12\left(3+\Lambda_2\right)P_2^2+ 12\left(3 - 2\Lambda_{12} +\Lambda_2-\Lambda_1\right)P_3^2  \\ 
\rm+ 24\left[ \left(3+ \Lambda_1+\Lambda_{12}\right)P_1+ \left(3 + \Lambda_{12}-\Lambda_2\right)P_2\right]P_3 \\
\rm +24\left(3-\Lambda_{12}\right)P_1 P_2-P_4^2-P_5^2- 24\left(V_{01}
e^{\xi_1}+V_{02} e^{\xi_2}\right),
\end{multline}
where $\rm \Lambda_1=\lambda_1^2/\triangle$, $\rm
\Lambda_2=\lambda_2^2/\triangle$, and $\rm
\Lambda_{12}=m^{12}\lambda_1 \lambda_2/\triangle$. Then, Hamilton's equations read
\begin{eqnarray} \label{ecs_mov_2}
\rm \dot \xi_1 &=& \rm 24 \left(3- \Lambda_1\right)P_1+24\left(3-\Lambda_{12} \right)P_2+ 24 \left(3+\Lambda_1+\Lambda_{12}\right)P_3 \nonumber\\
\rm \dot \xi_2 &=& \rm 24\left(3+ \Lambda_2\right)P_2 +24\left(3-\Lambda_{12}\right)P_1+ 24\left(3-   \Lambda_2+\Lambda_{12} \right)P_3, \nonumber\\
\rm \dot \xi_3 &=&\rm 24\left(3+\Lambda_1+\Lambda_{12}\right)P_1+24\left(3 - \Lambda_2+ \Lambda_{12}\right)P_2 +24 \left(3+\Lambda_2-\Lambda_1-2\Lambda_{12}\right)P_3 \nonumber\\
\rm \dot \xi_4 &=&\rm -2P_4 \,,\nonumber\\
\rm\dot \xi_5 &=&- \rm 2P_5 \,,\\
 \rm  \dot P_1 &=& \rm  24 V_{01}e^{\xi_1} \,,\nonumber\\
 \rm \dot P_2 &=& \rm 24V_{02}e^{\xi_2} \,, \nonumber\\
  \rm \dot P_3 &=& 0 \,, \nonumber\\
  \rm \dot P_4 &=& 0 \,, \nonumber\\
  \rm\dot P_5 &=& 0 \,.\nonumber
\end{eqnarray}
Here straightforwardly one can set $\rm P_i=p_i=constant$, with $\rm i=3,4,5$. Now we take the time derivative of $\rm \dot \xi_1$ (first
equation in (\ref{ecs_mov_2})), then we combine it with $\rm \dot{P}_1$, yielding
\begin{equation}\rm
\rm \ddot \xi_1= \rm 576V_{01} \left(3- \Lambda_1\right) e^{\xi_1}
+576V_{02}\left(3-\Lambda_{12}\right) e^{\xi_2} \,. \label{first}
\end{equation}
To find solutions of $(\Omega,\beta_+,\beta_-,\phi_1,\phi_2)$ we introduce the transformation (\ref{trans_2}) in order to separate the set of equations coming from the Hamiltonian density (\ref{hami-bi}), then we drop the mixed momenta by setting to zero their coefficients, therefore this procedure constraints the matrix element $\rm m^{12}$
\begin{equation}
\rm m^{12}=\frac{\lambda_1 \lambda_2}{6}\left[1 \pm \sqrt{1-
\left(\frac{6}{\lambda_1 \lambda_2}\right)^2}\right] \,,
\label{constraint}
\end{equation}
moreover, we fix the second term in the square root of (\ref{constraint}) to be a real number, and we consider $\lambda_1>0$, $\lambda_2>0$, hence
yielding the relation $\lambda_1\lambda_2 \geq 6$, which in turns ensures that $\rm m^{12}$ is always positive. Finally, the aforementioned simplifications yield the subsequent Hamilton equations
\begin{equation}\label{hamilton-equation}\small
\begin{split}
\rm \dot \xi_1&= \rm 24 \eta_1P_1 + 24 \left(9-\vert\eta_1\vert\right)p_3 \,,\\
\rm \dot \xi_2&= \rm 24\eta_2P_2 + 24\left(9-\eta_2 \right)p_3 \,, \\
\rm \dot \xi_3 &=\rm 24\left(9-\vert\eta_1\vert\right)P_1+
24\left(9 -\eta_2\right)P_2 +24 \left(-9+\vert\eta_1\vert+\eta_2\right)p_3 \,,\quad\\
\rm \dot \xi_4&=\rm -2p_4 \,,\\
\rm\dot\xi_5&\rm =-2p_5\\
\end{split}
\begin{split}
 \rm  \dot P_1 &= \rm  24 V_{01}e^{\xi_1} \,,\\
 \rm \dot P_2 &= \rm 24V_{02}e^{\xi_2} \,, \\
  \rm P_3 & = \rm p_3 \,, \\
  \rm P_4 &=  \rm p_4 \,, \\
  \rm P_5 &=  \rm p_5 \,, 
\end{split}
\end{equation}
with $\eta_1=3-\Lambda_1$ and $\eta_2=3+\Lambda_2$. In the next segments we will compute analytical expression provided distinct combinations of $\rm
(\lambda_1,\lambda_2)$. 
\subsection{Case $\rm \lambda_1\,\lambda_2=6$}
For this set of values we can see that $\rm m^{12}=1$, and $\triangle=2$. We also set $\rm \lambda_1=6/\lambda_2$, and restrict our results by fixing $\lambda_2\not=\sqrt{6}$. We start taking the time derivative of $\rm \dot \xi_1$ (from (\ref{hamilton-equation})), so we have a differential equation for the variable
$\xi_1$,
\begin{equation}
\rm \ddot \xi_1= 576\eta_1 \,V_{01} e^{\xi_1} \,, \label{xi-1}
\end{equation}
which its solution has the form
\begin{equation}
\rm e^{\xi_1}=\frac{ r_1^2}{288 \vert\eta_1\vert V_{01}}\left\{
\begin{tabular}{ll}
$\rm  Sech^2\left(r_1 t-q_1\right)$\,, & $\rm \lambda_1 > \sqrt{6}$
\quad\mbox{at}\quad $\eta_1<0$ \,. \\
$\rm  Csch^2\left(r_1 t-q_1\right)$\,, & $\rm \lambda_1 < \sqrt{6}$
\quad\mbox{at}\quad $\eta_1>0$ \,. \label{solucion-xi1}
\end{tabular}
\right.
\end{equation}
Note that this solution depends strongly on the value of $\rm \lambda_1$. Moreover, $\rm \dot\xi_2$ has the
same functional structure as $\rm\dot\xi_1$ when $\rm\eta_1>0$, since
$\rm\eta_2>0$ for all $\rm\lambda_2$, therefore its solution is
\begin{equation}
\rm e^{\xi_2}=\frac{ r_2^2}{288\eta_2\,V_{02}} \, Csch^2\left(r_2
t-q_2\right), \label{solucion-xi2}
\end{equation}
where $\rm r_i$ and $\rm q_i$ (with $\rm i=1,2$) are integration constants of both solutions (\ref{solucion-xi1}) and (\ref{solucion-xi2}). Given that two distinct solutions emerge due to $\rm\lambda_1$ there are indeed two different scenarios: phantom and quintessence. Thus we will analyse both cases. 
\subsubsection{Phantom domination: solution when
$\rm \lambda_1>\sqrt{6}\,\, (\eta_1<0)$, and $\rm\lambda_2<\sqrt{6}$. }\label{phantom_dom}
We start with the solutions
\begin{eqnarray}
\rm e^{\xi_1}&=&\rm  \frac{
r_1^2}{288 \vert\eta_1\vert V_{01}}\,Sech^2\left(r_1
t-q_1\right) \,,\nonumber\\
\rm e^{\xi_2}&=&\rm\frac{ r_2^2}{288\eta_2\,V_{02}} \, Csch^2\left(r_2
t-q_2\right) \,,
\end{eqnarray}
then we substitute them into Hamilton equations for the momenta (\ref{hamilton-equation}), obtaining
\begin{eqnarray}
\rm P_1 &=& \rm p_1 + \frac{r_1}{12 \vert\eta_1\vert } \, Tanh\left(r_1 t-q_1
\right), \label{solution-p1-1} \\
\rm P_2 &=& \rm p_2 - \frac{r_2}{12\eta_2} \, Coth\left(r_2 t-q_2
\right), \label{solution-p2-1}
\end{eqnarray}
here $\rm p_i$ ($\rm i=1,2$) are integration constants. It can be easily verified that the Hamiltonian is identically zero when
\begin{eqnarray}
\rm p_1=\frac{ \vert\eta_1\vert +9}{ \vert\eta_1\vert } p_3 , \qquad 
p_2=\frac{\eta_2-9}{\eta_2} p_3,\nonumber\\
\rm p_3= +\frac{1}{36}\sqrt{\frac{\eta_2 r_1^2- \vert\eta_1\vert  r_2^2+12 \vert\eta_1\vert \eta_2(p_4^2+p_5^2) }{3\left[ \vert\eta_1\vert \eta_2-3 \vert\eta_1\vert +3\eta_2 \right]}} \,. \label{coefficients}
\end{eqnarray}
Thus the solutions for the $\rm \xi_i$ coordinates become
\begin{eqnarray}
\rm \xi_1&=& \rm Ln\left(\frac{r_1^2}{288 \vert\eta_1\vert V_{01}}\right)+Ln\left[ Sech^2\left(r_1\,t -q_1 \right)\right] \,,  \label{xi1-1}\\
\xi_2 &=& \rm Ln\left(\frac{r_2^2}{288\eta_2\, V_{02}}\right)+Ln\left[Csch^2\left(r_2\,t -q_2 \right)\right] \label{xi2-1} \\
\rm \xi_3&=& \rm a_3 + 648\frac{ \vert\eta_1\vert \eta_2-3 \vert\eta_1\vert +3\eta_2}{ \vert\eta_1\vert \eta_2} p_3 t+\frac{9+ \vert\eta_1\vert }{ \vert\eta_1\vert }Ln\left[Cosh^2\left(r_1\,t - q_1 \right)\right] \\
& &+\rm \frac{\eta_2-9}{\eta_2}Ln\left[Sinh^2\left(r_2\,t - q_2\right) \right],\nonumber\\
\rm \xi_4 &=& \rm a_4 -2p_4 t \,, \\
\rm \xi_5 &=& \rm a_5 -2p_5 t \,,
\end{eqnarray}
here $\rm a_i$ ($ \rm i=3,4,5$) stand as constants coming from integration. After applying the inverse canonical transformation we get the solutions in terms of the original variables $\rm (\Omega, \phi_1, \phi_2,\beta_+,\beta_-)$,
\begin{eqnarray}\label{sols_quintessence}\small
\rm \Omega &=& \rm  \Omega_0 + Ln\left[ Cosh^{\beta_1} \left(r_1\,t-q_1 \right) Csch^{\beta_2}\left(r_2\, t -q_2 \right)\right]\rm +36\frac{ \vert\eta_1\vert \, \eta_2-3 \vert\eta_1\vert + 3\eta_2}{ \vert\eta_1\vert \eta_2}p_3 t \,,\nonumber \\
\rm \phi_1 &=& \rm \phi_{10}+ Ln\left[Cosh^{\frac{2\left( \vert\eta_1\vert +3 \right)}{\lambda_1 \vert\eta_1\vert}}\left(r_1\,t-q_1\right)Csch^{\frac{6}{\lambda_1 \eta_2}}\left(r_2\, t
-q_2\right)\right] \rm +216\frac{\vert\eta_1\vert \eta_2-3 \vert\eta_1\vert +3\eta_2}{\lambda_1\,\vert\eta_1\vert \eta_2} p_3 t \,,\nonumber \\
 \rm \phi_2 &=& \rm \phi_{20}+Ln\left[Cosh^{\frac{6}{\lambda_2 \vert\eta_1\vert}}\left(r_1\,t-q_1\right)Sinh^{\frac{2\left( \eta_2-3\right)}{\lambda_2 \eta_2}}\left(r_2\, t
-q_2\right)\right] \rm +216\frac{ \vert\eta_1\vert  \eta_2-3 \vert\eta_1\vert +
3\eta_2}{\lambda_2\, \vert\eta_1\vert \eta_2} p_3 t \,,\nonumber\\
\rm \beta_+ &=& \rm a_4 -2p_4 t,\nonumber\\
\rm \beta_- &=& \rm a_5 -2p_5 t,
\end{eqnarray}
where $\rm \beta_1=1/ \vert\eta_1\vert$, $\rm \beta_2=1/\eta_2$, and the constants $\rm \Omega_0, \phi_{10}$, and $\rm \phi_{20}$ are given by
\begin{eqnarray}
&&\rm \Omega_0=\rm Ln\left[\frac{r_1\,r_2}{288\sqrt{ \vert\eta_1\vert \eta_2
V_{01}\,V_{02}}} \right]^{\frac{1}{9}}+\frac{a_3}{18} \,,\nonumber\\
&&\rm \phi_{10}= \rm Ln\left[\frac{12\sqrt{2}r_2\vert\eta_1\vert V_{01}}{r_1^2\sqrt{\eta_2 V_{02}}}\right]^{\frac{2}{3\lambda_1}}+\frac{a_3}{3\lambda_1} \,, \\
&&\rm \phi_{20}=\rm Ln\left[\frac{12\sqrt{2}r_1\eta_2 V_{02}}{r_2^2\sqrt{ \vert\eta_1\vert V_{01}}}\right]^{\frac{2}{3\lambda_2}}+\frac{a_3}{3\lambda_2} \,. \nonumber 
\end{eqnarray}
Therefore the scale factors are
\begin{eqnarray}
\rm A(t)&=&\rm \left[\frac{r_1 r_2}{288\sqrt{\vert\eta_1\vert\eta_2 V_{01}V_{02}}} \right]^{\frac{1}{9}}\,e^{\frac{a_3+a_4+\sqrt{3}a_5}{18}}Cosh^{\beta_1} \left(r_1\,t -q_1 \right) Csch^{\beta_2}\left(r_2\, t-q_2 \right)\,\nonumber\\
&&\rm \times Exp\left\{\left[36\frac{\vert\eta_1\vert\, \eta_2-3 \vert\eta_1\vert +
3\eta_2}{\vert\eta_1\vert \eta_2} p_3-2p_4-2\sqrt{3}p_5\right]
t\right\} \,,\label{scale-a}\\
\rm B(t)&=&\rm \left[\frac{r_1 r_2}{288\sqrt{ \vert\eta_1\vert \eta_2 V_{01}
V_{02}}} \right]^{\frac{1}{9}}\,e^{\frac{a_3+a_4-\sqrt{3}a_5}{18}}
Cosh^{\beta_1} \left(r_1\,t -q_1 \right) Csch^{\beta_2}\left(r_2\, t
-q_2 \right)\, \nonumber\\
&&\rm \times Exp\left\{\left[36\frac{ \vert\eta_1\vert \eta_2-3 \vert\eta_1\vert +
3\eta_2}{ \vert\eta_1\vert\eta_2} p_3-2p_4+2\sqrt{3}p_5\right]
t\right\} \,,\label{scale-b}\\
\rm C(t)&=&\rm \left[\frac{r_1 r_2}{288\sqrt{\vert\eta_1\vert\eta_2 V_{01}
V_{02}}} \right]^{\frac{1}{9}} e^{\frac{a_3-2a_4}{18}} Cosh^{\beta_1}
\left(r_1\,t -q_1 \right) Csch^{\beta_2}\left(r_2\, t -q_2
\right)\,\nonumber\\
&&\rm \times Exp\left\{\left[36\frac{\vert\eta_1\vert \eta_2-3\vert\eta_1\vert +
3\eta_2}{ \vert\eta_1\vert \eta_2} p_3+4p_4\right] t\right\} \,,\label{scale-c}
\end{eqnarray}
and the volume function $\rm V(t)=ABC=e^{3\Omega}$ becomes
\begin{equation}
\rm V(t)= \left[\frac{r_1 r_2}{288\sqrt{ \vert\eta_1\vert\eta_2 V_{01} V_{02}}}
\right]^{\frac{1}{3}}\,e^{\frac{a_3}{6}} Cosh^{3\beta_1}
\left(r_1\,t -q_1 \right) Csch^{3\beta_2}\left(r_2\, t -q_2 \right)
\rm  Exp\left\{108\frac{\vert\eta_1\vert \eta_2-3 \vert\eta_1\vert +
3\eta_2}{ \vert\eta_1\vert \eta_2} p_3 t\right\} \,.\label{volum3}
\end{equation}
\begin{figure}[ht!]
\begin{center}
\includegraphics[scale=0.4]{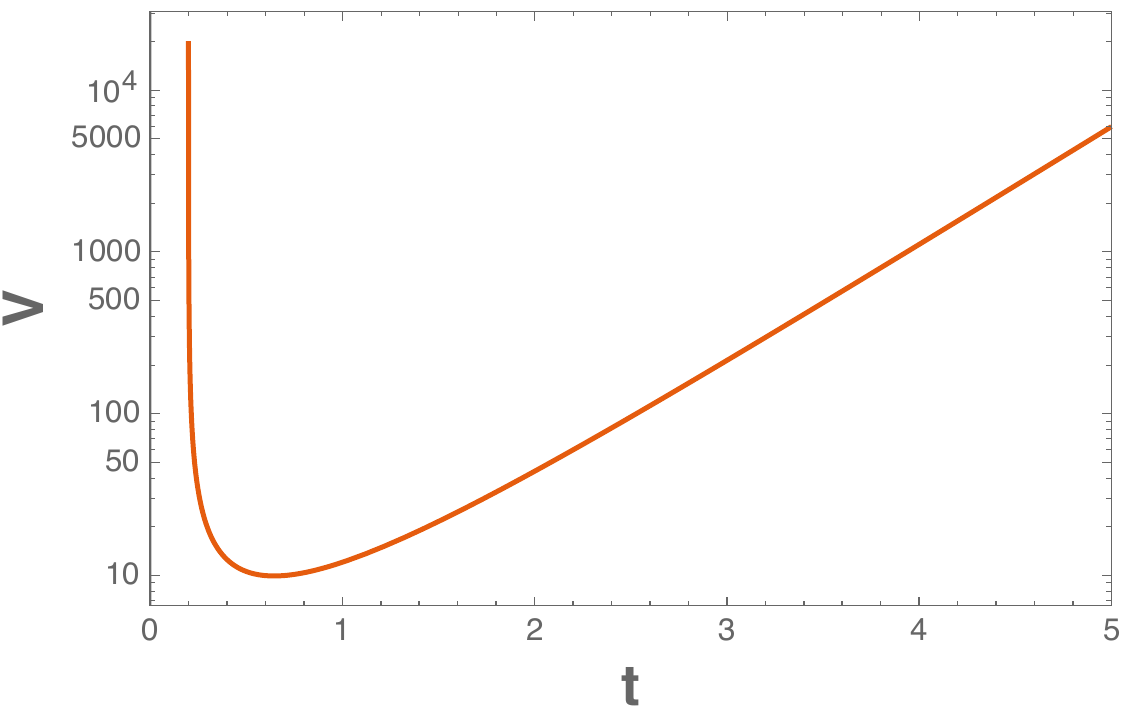}
\includegraphics[scale=0.4]{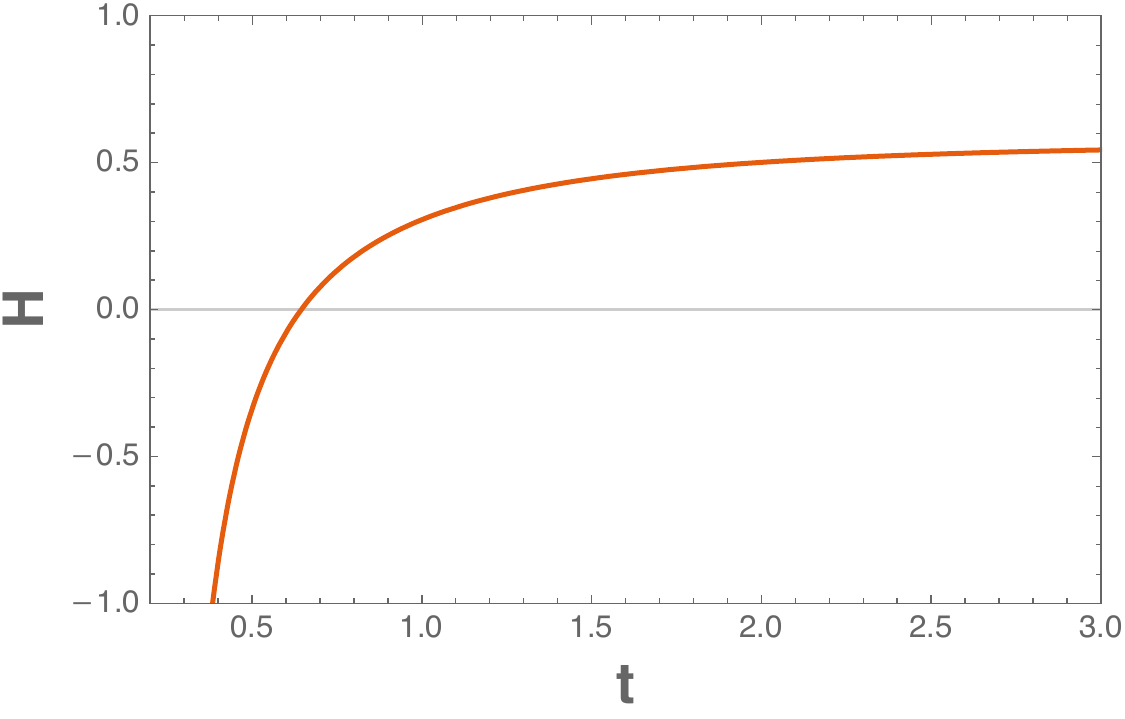}
\caption{This figure shows the time evolution of the volume function $\rm V=V(t)=ABC$ and the Hubble parameter $\rm H=H(t)$. We use arbitrary units of $\rm V_{01} = 5.0 \,, V_{02} = 10^{-5}$, $ a_3=-0.5$, $\rm q_1=q_2=0.1$, $\rm r_1=16$, $\rm r_2=0.5$ and $\rm
\lambda_1=4.3\sqrt{6}$, $\rm p_4=p_5=0.001$. Recall that $\rm
\lambda_2=\sqrt{6}/\lambda_1$, and other constants depend on
the aforementioned values.}\label{figura_1}
\end{center}
\end{figure}

In Fig.~\ref{figura_1} we can appreciate the evolution of the volume function $\rm V$ and the Hubble parameter $\rm H$, with respect to time. Note that this multi-field cosmological framework avoids the ``big-bang'' singularity by means of a ``big-bounce'', and this behaviour is also evident due to the horizontal crossing of the Hubble parameter (at $\rm H=0$) where in fact corresponds at the time of the ``big-bounce''. Indeed, this outcome has been already pointed out in \cite{Socorro:2022aoz}; however, authors studied a FLRW framework. 

In this subsection there is another case left, when $\lambda_1<\sqrt{6}$ and $\lambda_2>\sqrt{6}$ (which corresponds to $\rm \eta_1>0$); however, we have not included this scenario since the volume function decreases in time, hence becoming physically unfeasible. Nonetheless, we will continue examining the instance where $\lambda_1=\lambda_2=\sqrt 6$.

\subsection{Case $\lambda_1=\lambda_2=\sqrt{6}$}
\begin{figure}[ht!]
\begin{center}
\includegraphics[scale=0.4]{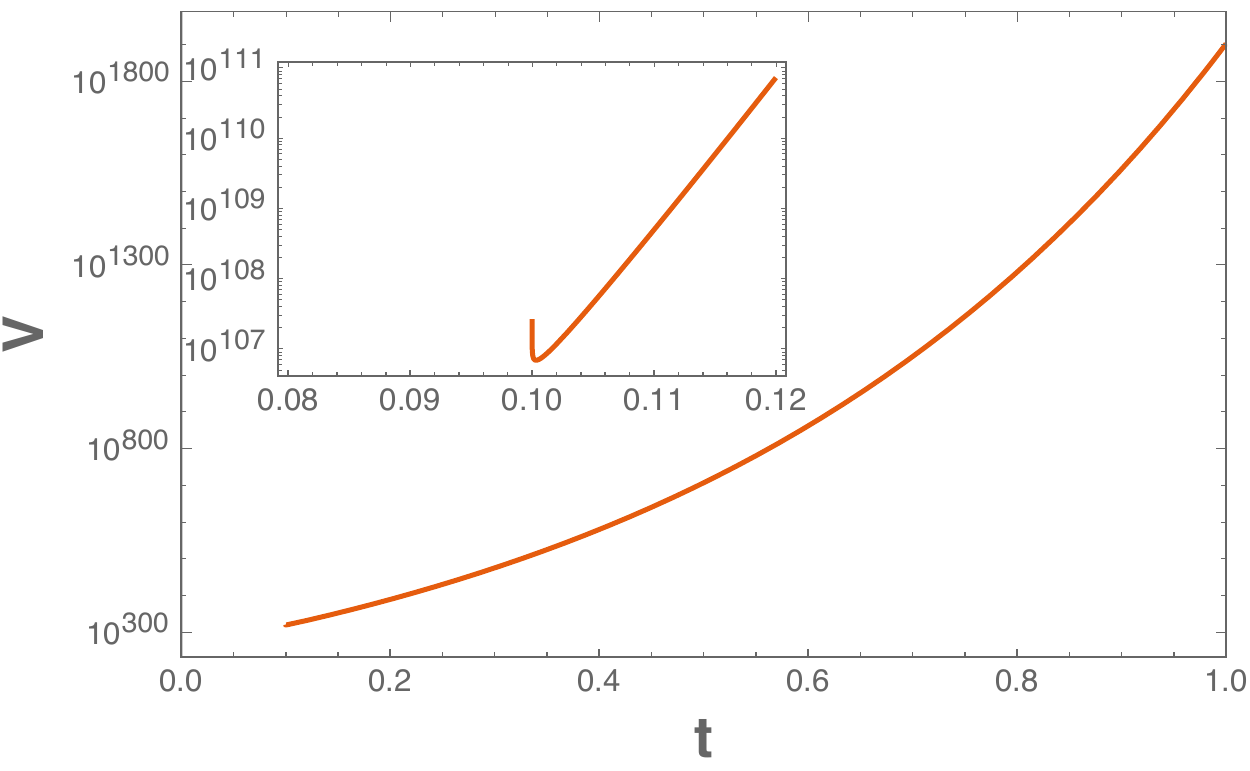}
\includegraphics[scale=0.4]{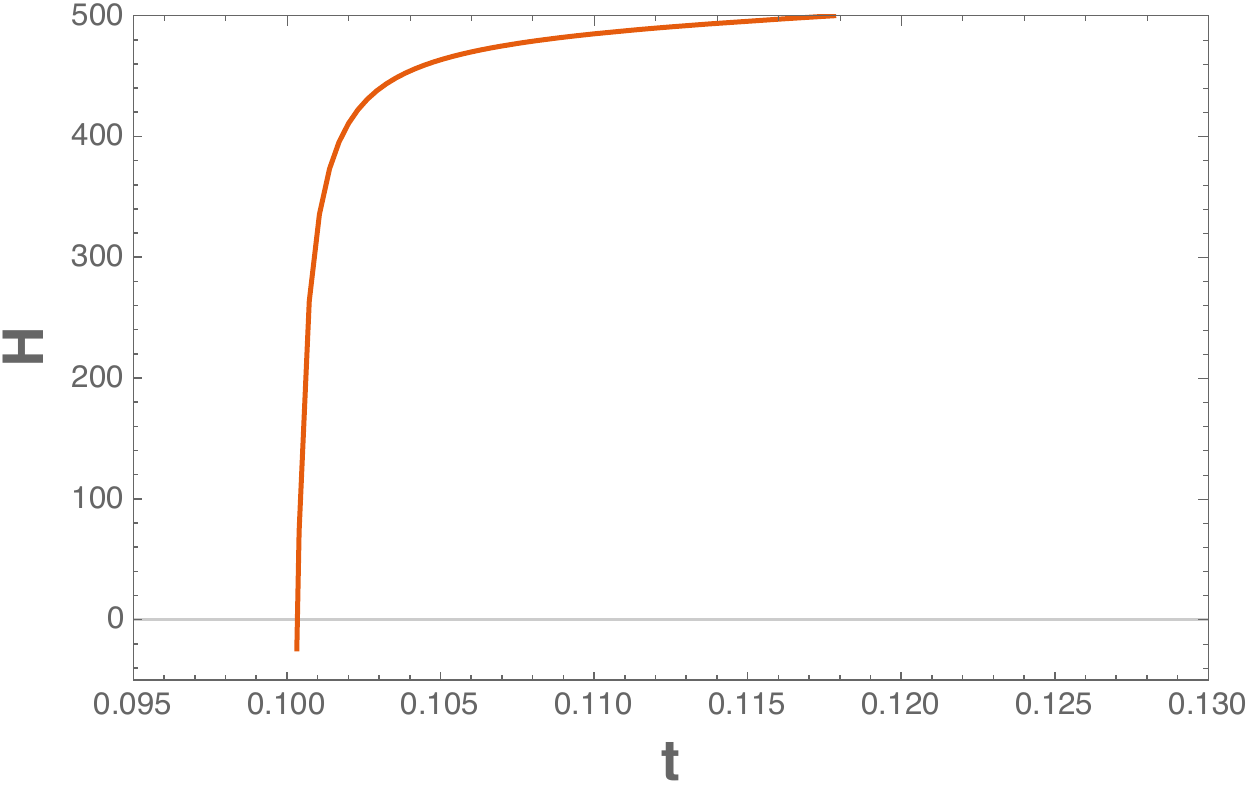}
\caption{This figure shows the time evolution of the volume function V = ABC and the Hubble parameter H(t). We use arbitrary units, namely $\rm
\lambda_1=\lambda_2=\sqrt{6}$, $\rm V_{01} = 1$, $\rm V_{02} = 0.1$, $\rm
a_1=a_3=a_4=a_5=1$, $\phi_1=\beta_+=\beta_-=1$ and $\Psi_0=1$.}\label{figura_2}
\end{center}
\end{figure}

For these particular values we have that $\Lambda_1=\Lambda_2=3$ with $\eta_1=0$ and $\eta_2=6$, then the Hamilton equations reduce to
\begin{equation}\label{caso-igual}
\begin{split}
\rm \dot \xi_1&= \rm  216 p_3 \,,\\ 
\rm \dot \xi_2&= \rm 144P_2 + 72p_3 \,, \\
\rm \dot \xi_3 &=\rm 216P_1+ 72P_2 -72p_3 \,,\qquad\\
\rm \dot \xi_4&=\rm -2p_4 \,,\\ 
\rm \dot \xi_5&=\rm -2p_5 \,,\\ 
\end{split}
\begin{split}
\rm  \dot P_1 &= \rm  24 V_{01}e^{\xi_1} \,,\\
\rm \dot P_2 &= \rm 24V_{02}e^{\xi_2} \,, \\
\rm P_3 &= \rm p_3\,,\, P_4= p_4 \,,\, P_5= p_5 \,.
\end{split}
\end{equation}
Right away from $\rm \dot \xi_1$, we have its solution
\begin{equation}
\rm\xi_1= a_1+216p_3 t \,,
\end{equation}
here $\rm a_1$ is an integration constant. Then, taking the time derivative of $\dot\xi_2$ results in $\rm \ddot \xi_2=3456\, V_{02}\,e^{\xi_2}$, having a solution of the form
\begin{equation}\rm
\xi_2=Ln\left(\frac{\alpha_2^2}{1728\, V_{02}}
\right)+Ln\left[Csch^2(\alpha_2\,t-\beta_2) \right] \,.
\end{equation}
Now we know the functional form of $\rm \xi_1$ and $\rm \xi_2$, then we can compute the remaining momenta
\begin{eqnarray}
\rm P_1(t)&=&\rm p_1+\frac{V_{01}}{9p_3}\, e^{a_1+216p_3\,t}, \nonumber\\
\rm P_2(t)&=&\rm p_2-\frac{\alpha_2}{72}\, Coth(\alpha_2\,t -
\beta_2) \,.
\end{eqnarray}
And for the rest of the variables $\rm \xi_j$ we have
\begin{eqnarray}\rm
\xi_3 &=&\rm
a_3+(216p_1-108p_3)t+\frac{V_{01}}{9p_3^2}e^{a_1+216p_3t}+Ln\left[Csch\left(\alpha_2\,t-\beta_2
\right) \right] \,, \nonumber\\
\rm  \xi_4&=&\rm a_4 - 2p_4 t \,,\nonumber\\
\rm  \xi_5&=&\rm a_5 - 2p_5 t \,, \label{chis}
\end{eqnarray}
where $\rm a_3,\, a_4,\, a_5$ are integration constants. By reinserting these momenta into the Hamiltonian density (\ref{new-moment}) leads to the constraints
\begin{equation}\rm
p_2=-\frac{p_3}{2} \,,\quad 
p_3=2p_1 \pm \frac{\sqrt{3}}{108}\sqrt{\alpha_2^2+15552\,p_1^2-72(p_4^2+p_5^2)} \,.
\end{equation}
We then go back to our original variables $(\Omega,\beta_+,\beta_-,\phi_1,\phi_2)$, hence we have 
\begin{eqnarray}
\rm \Omega(t)&=&\rm \frac{a_1+a_3}{18}+Ln\left[\frac{\alpha_2}{24\sqrt{3\,V_{02}}} \right]^{\frac{1}{9}} \rm +Ln\left[Csch^{\frac{1}{6}}\left(\alpha_2\,t-\beta_2 \right)\right]+(12p_1+6p_3)t+\frac{V_{01}}{162p_3^2}e^{a_1+216p_3t},\label{omega}\nonumber\\
\rm \phi_1(t)&=&\rm\frac{-2a_1+a_3}{3\lambda_1}+Ln\left(\frac{\alpha_2^2}{1728\, V_{02}}\right)^{\frac{1}{3\lambda_1}}\rm +\frac{1}{\lambda_1}\left[\left(72p_1-180p_3\right)t+\frac{V_{01}}{27p_3^2}e^{a_1+216p_3t}+Ln\left[Csch\left(\alpha_2\,t-\beta_2 \right) \right]\right], \nonumber \label{phi1}\\
\rm \phi_2(t)&=&\rm\frac{a_1+a_3}{3\lambda_2}+Ln\left(\frac{\alpha_2^2}{1728\, V_{02}}\right)^{-\frac{2}{3\lambda_2}}
\rm + \frac{1}{\lambda_2}\left[\left(72p_1+36p_3\right)t+ \frac{V_{01}}{27p_3^2}\,e^{a_1+216p_3t}+Ln\left[Sinh\left(\alpha_2\,t-\beta_2 \right)
\right] \right],\nonumber \label{phi12}\\
\rm \beta_+&=& \rm a_4 -2p_4 t, \nonumber\\
\rm \beta_-&=& \rm a_5 -2p_5 t \,,
\end{eqnarray}
therefore the scale factors become
\begin{eqnarray}
\rm A&=& \rm \left[\frac{\alpha_2}{24\sqrt{3\,V_{02}}}
\right]^{\frac{1}{9}}e^{\frac{a_1+a_3+18a_4+18\sqrt{3}a_5}{18}}Csch^{\frac{1}{6}}\left(\alpha_2\,t-\beta_2\right) 
\rm \times Exp\left[2(6p_1+3p_3-p_4-\sqrt{3}p_5)t+\frac{V_{01}}{162p_3^2}e^{a_1+216p_3t} \right] \,,\nonumber\\
\rm B&=&\rm \left[\frac{\alpha_2}{24\sqrt{3\,V_{02}}}
\right]^{\frac{1}{9}}e^{\frac{a_1+a_3+18a_4-18\sqrt{3}a_5}{18}}\,Csch^{\frac{1}{6}}\left(\alpha_2\,t-\beta_2\right)\rm \times Exp\left[2(6p_1+3p_3-p_4+\sqrt{3}p_5)t
+\frac{V_{01}}{162p_3^2}e^{a_1+216p_3t} \right] \nonumber,\\
\rm C&=&\rm \left[\frac{\alpha_2}{24\sqrt{3\,V_{02}}}\right]^{\frac{1}{9}}e^{\frac{a_1+a_3-36a_4}{18}}\,Csch^{\frac{1}{6}}\left(\alpha_2\,t-\beta_2\right) \rm \rm Exp\left[2(6p_1+3p_3+2p_4)t+\frac{V_{01}}{162p_3^2}e^{a_1+216p_3t} \right] \,,
\end{eqnarray}
and the volume function $\rm V(t)=ABC$ is
\begin{equation}
\rm V(t)=\left[\frac{\alpha_2}{24\sqrt{3\,V_{02}}}\right]^{\frac{1}{3}}e^{\frac{a_1+a_3}{6}}\rm Csch^{\frac{1}{2}}\left(\alpha_2\,t-\beta_2\right) \rm \times Exp\left[18(2p_1+p_3)t +\frac{V_{01}}{54p_3^2}e^{a_1+216p_3t}\right]
\end{equation}
Fig.~\ref{figura_2} shows the time evolution of the volume function $\rm V$ and the Hubble parameter $\rm H$. At first glance $\rm V$ exhibits only substantial growth; nonetheless, when zooming in a small bounce can be appreciated at a very short time scale. Thus $\rm V$ circumvents again the ``big-bang'' singularity. Indeed, this bounce is more perceptible on the dynamical evolution of $\rm H$, happening at the time $\rm H=0$.   
We can measure the anisotropic density implementing the Misner's parameterization
\begin{eqnarray}
\rm a &=& \rm  \Omega+\beta_+\sqrt{3} \beta_- \,, \qquad b= \rm \Omega+\beta_+-\sqrt{3} \beta_- \,,\qquad c= \rm \Omega-2\beta_+ \,,\nonumber\\
\rm \Omega&=& \rm \frac{1}{3}(a+b+c) \,,\qquad  \beta_+=\frac{1}{6}(a+b-2c) \,, \qquad\quad \beta_-=\frac{\sqrt{3}}{6}(a-b) \,.
\end{eqnarray}
The anisotropic and gravitational densities are defined by $\rm\rho_{anisotropic}=(\dot \beta_+)^2 + (\dot \beta_-)^2$ and $\rho_\Omega=(\dot \Omega)^2$, respectively. When the anisotropic-to-gravitational density rate goes to zero ($\rm\rho_{anisotropic}/\rho_\Omega \rightarrow 0 $) the spacetime becomes isotropic \cite{Socorro:2014ama}. Remarkably, in all cases the anisotropic density is 
\begin{equation}
\rm \rho_{anisotropic}= 4(p_4^2+ p_5^2)=constant \,,    
\end{equation}
and since $\rm\dot{\Omega}$ increases with time, isotropization is indeed reached eventually. 

%
\section{Quantum scheme}
%
The quantum scheme is obtained by making the usual replacement $\rm \Pi_{q^\mu}=-i\hbar\partial_{q^\mu}$ into the classical Hamiltonian density. Also, in order to consider different factor orderings among $\rm e^{-3\Omega}$ and $\rm \Pi_\Omega$, we take $\rm e^{-3\Omega}\Pi_\Omega^2 \to
e^{-3\Omega}\left[\Pi_\Omega^2 +Qi\hbar \Pi_\Omega \right]$ where $\rm Q$
is a real number that measures the ambiguity in the factor ordering. We therefore write the Hamiltonian (\ref{hamiltonian})
\begin{equation} 
\rm {\cal H}=\Pi_\Omega^2 +Qi\hbar \Pi_\Omega-\Pi_+^2-\Pi_-^2-12\Lambda_2\Pi_{\phi_1}^2+12\Lambda_1 \Pi_{\phi_2}^2
\rm -24 \Lambda_0\Pi_{\phi_1}\Pi_{\phi_2} -24V_{01} e^{-\lambda_1\phi_1+6\Omega} -24V_{02}e^{-\lambda_2\phi_2+6\Omega} \,, \label{q-hamifrw-mod}
\end{equation}
where we have taken into account the constraint (\ref{constraint})
of the matrix element $\rm m^{12}$, and we have fixed the gauge to
$\rm N=24 e^{3\Omega}$. Once again we consider the canonical transformation (\ref{trans_2}) $\rm(\Omega,\phi_1,\phi_2,\beta_+,\beta_-)\leftrightarrow
(\xi_1,\xi_2,\xi_3,\xi_4,\xi_5)$, and the new momenta (\ref{new-moment}); hence we end up with 
\begin{multline}
\rm {\cal H} = \rm  12 \eta_1 P_1^2+12\eta_2P_2^2+ 12\left(-9 +\eta_1+\eta_2\right)P_3^2 + 24P_3\left[ \left(9-\eta_1 \right)P_1+\left(9 -\eta_2\right)P_2\right]\\\rm +6Qi\hbar \left(P_1+P_2+P_3\right)\rm - (P_4^2 +P_5^2)- 24\left(V_{01} e^{\xi_1}+V_{02}
e^{\xi_2}\right),\label{q-hamiltonian}
\end{multline}
here $\eta_1=3-\Lambda_1$ and $\eta_2=3+\Lambda_2$. Therefore the corresponding quantum Hamiltonian operator becomes
\begin{multline}
\rm \hat{\cal H}\Psi(\xi_i) = \rm  -12\hbar^2 \eta_1
\frac{\partial^2 \Psi}{\partial \xi_1^2}-12\hbar^2
\eta_2\frac{\partial^2 \Psi}{\partial \xi_2^2}- 12\hbar^2\left(-9 +
\eta_1+\eta_2\right)\frac{\partial^2 \Psi}{\partial \xi_3^2}
\\
-\hbar^2 24\left[ \left(9-\eta_1 \right)\frac{\partial^2
\Psi}{\partial \xi_1 \partial \xi_3}+ \left(9
-\eta_2\right)\frac{\partial^2 \Psi}{\partial \xi_2 \partial
\xi_3}\right]\rm+6Q\hbar^2 \left(\frac{\partial
\Psi}{\partial \xi_1} +\frac{\partial \Psi}{\partial
\xi_2}+\frac{\partial \Psi}{\partial \xi_3}\right) \\
+\hbar^2\left(\frac{\partial^2 \Psi}{\partial \xi_4^2}+\frac{\partial^2\Psi}{\partial \xi_5^2}\right)- 24\left(V_{01} e^{\xi_1}+V_{02}e^{\xi_2}\right)\Psi=0 \,.\label{wdw}
\end{multline}
Note that expression~(\ref{wdw}) is the Wheeler-DeWitt (WDW) equation. To find the wave function we propose the following ansatz $\rm
\Psi=e^{(p_3 \xi_3+p_4 \xi_4+p_5 \xi_5)}{\cal B}(\xi_1,\xi_2)$ with
$\rm p_j=constant, j=3,4,5$; then the WDW equation can be separated as
%
\begin{equation}
\begin{split}
\rm{\cal H}\Psi &= \rm  -12\hbar^2 \eta_1 \frac{\partial^2 {\cal B}}{\partial \xi_1^2}+6\hbar^2\left(Q
-4p_3(9-\eta_1)\right)\frac{\partial {\cal B}}{\partial\xi_1}\rm +3\hbar^2\,\left[p_3\left(Q-2p_3(9-\eta_1+\eta_2)\right)+\frac{(p_4^2 +p_5^2)}{6}-8\frac{V_{01}}{\hbar^2}e^{\xi_1}\right]{\cal B} \\ 
& \rm-12\hbar^2 \eta_2 \frac{\partial^2 {\cal B}}{\partial
\xi_2^2}+6\hbar^2\left(Q -4p_3(9-\eta_2)\right)\frac{\partial {\cal
B}}{\partial \xi_2}\rm +3\hbar^2\,\left[p_3\left(Q-2p_3(9-\eta_1+\eta_2)
\right)+\frac{(p_4^2 +p_5^2)}{6}-8\frac{V_{02}}{\hbar^2}
e^{\xi_2}\right]{\cal B}=0 \,. \label{quantum}
\end{split}
\end{equation}
%
Additionally we assume that $\rm {\cal B}(\xi_1,\xi_2)={\cal B}_1(\xi_1){\cal B}_2(\xi_2)$, having;
\begin{equation}
\rm -12 \eta_1 \frac{d^2 {\cal B}_1}{d \xi_1^2}+6\left(Q-4p_3(9-\eta_1)\right)\frac{d {\cal B}_1}{d\xi_1}\rm +3\,\left[-\nu^2+p_3\left(Q-2p_3(9-\eta_1+\eta_2)
\right)+\frac{(p_4^2 +p_5^2)}{6}-8\frac{V_{01}}{\hbar^2}e^{\xi_1}\right]{\cal B}_1=0
\end{equation}
\begin{equation}
\rm -12 \eta_2 \frac{d^2 {\cal B}_2}{d \xi_2^2}+6\left(Q-4p_3(9-\eta_2)\right)\frac{d {\cal B}_2}{d\xi_2}\rm+3\,\left[\nu^2+p_3\left(Q-2p_3(9-\eta_1+\eta_2)\right)+\frac{(p_4^2 +p_5^2)}{6}-8\frac{V_{02}}{\hbar^2}e^{\xi_2}\right]{\cal B}_2=0 \,,
\end{equation}
where for convenience we have written the separation constant as $\rm 3\nu^2$. In the following sections we will show that quantum solutions can be divided into two classes, which will depend on combinations of $\lambda_1,\lambda_2$. Moreover, in~\cite{Hartle:1983ai,Hawking:1983hj} it has been shown that the best candidates for quantum solutions are wave functions that have a damping behavior with respect to the scale factor, since these allow to obtain good classical solutions when using the WKB approximation for any scenario in the evolution of our universe. Thus, the quantum analytical solutions to be presented in this paper will have this characteristic, featuring a damping behavior with respect to the average scale factor (cubic root of the isotropic volume $\rm V$). 

\subsection{Quantum solution for
$\lambda_1>\sqrt{6}\, \left(\eta_1<0\right)$, and $\lambda_2<\sqrt{6}$.}

We have to find solutions of ${\cal B}_1(\xi_1)$ and ${\cal B}_2(\xi_2)$ given the conditions $\lambda_1>\sqrt{6}\, \left(\eta_1<0\right)$, and $\lambda_2<\sqrt{6}$; however, we present only the explicit procedure for finding ${\cal B}_1(\xi_1)$. The same method is applied in order to obtain $\rm {\cal B}_2$. Hence, the equation for $\rm {\cal B}_1$ is
\begin{equation} 
\rm \frac{d^2 {\cal B}_1}{d \xi_1^2}+\frac{\left(Q-4p_3(9+\vert\eta_1\vert)\right)}{2\vert\eta_1\vert}\frac{d {\cal B}_1}{d
\xi_1}
\rm +\frac{1}{4\vert\eta_1\vert}\left[-\nu^2+p_3\left(Q-2p_3(9+\vert\eta_1\vert+\eta_2)\right)+\frac{(p_4^2 +p_5^2)}{6}-8\frac{V_{01}}{\hbar^2}e^{\xi_1}\right]{\cal B}_1=0.
\end{equation}
In fact this equation can be written as $\rm y^{\prime \prime} + a
y^\prime + \left(b e^{\kappa x } +c \right)y=0$, and its solution is of the form~\cite{polyanin}
\begin{equation}
\rm Y(x)=Exp\left({-\frac{ax}{2}}\right)
Z_\rho\left(\frac{2\sqrt{b}}{\kappa} e^{\frac{\kappa x}{2}}\right),
\end{equation}
here $\rm Z_\rho$ are generic Bessel function with order $\rm
\rho=\sqrt{a^2-4c}/\kappa$. If $\sqrt{b}$ is real, $\rm Z_\rho$ are
the ordinary Bessel function, otherwise the solution will be given
by the modified Bessel function. Then we have the following relations
\begin{eqnarray}
\rm a&=&\rm\frac{Q-4p_3(9+\vert\eta_1\vert)}{2\vert\eta_1\vert} \,,\nonumber\\
\rm b&=&\rm -\frac{2V_{01}}{\hbar^2\vert\eta_1\vert} \nonumber,\\
\rm c&=&\rm \frac{1}{4\vert\eta_1\vert}\left[-\nu^2+p_3\left(Q-2p_3(9+\vert\eta_1\vert+\eta_2)\right)+\frac{(p_4^2 +p_5^2)}{6}\right] \,,\\
\rm\kappa&=&1, \nonumber
\end{eqnarray}
since $\sqrt{b}$ is imaginary, then the function $\rm Z_\rho$ becomes the
modified Bessel function $\rm K_\rho$. Thus, ${\cal B}_1$ is
\begin{equation}
\rm {\cal B}_1=Exp\left[-\frac{\left(Q
-4p_3(9+\vert\eta_1\vert)\right)}{4\vert\eta_1\vert}\xi_1\right]\,
K_{\rho_1}\left[\frac{2}{\hbar}\sqrt{\frac{2V_{01}}{\vert\eta_1\vert}}
e^{\frac{\xi_1}{2}}\right] \,, \label{b1}
\end{equation}
with 
\begin{equation}\small
\rm\rho_1=\sqrt{\left[\frac{\left(Q
-4p_3(9+\vert\eta_1\vert)\right)}{2\vert\eta_1\vert}\right]^2-\frac{1}{\vert\eta_1\vert}\left[-\nu^2+p_3\left(Q-2p_3(9+\vert\eta_1\vert+\eta_2)
\right)+\frac{(p_4^2 +p_5^2)}{6}\right]} \,.    
\end{equation}
Then for the solution of $\rm {\cal B}_2$ we have the following relations
\begin{eqnarray}
  \rm a&=& \rm -\frac{6\left(Q-4p_3(9-\eta_2)\right)}{2\eta_2},\nonumber\\ 
  \rm b&=&\rm\frac{2V_{02}}{\hbar^2\eta_2},\nonumber \\
\rm c&=&-\rm \frac{1}{4\eta_2}\left[\nu^2+p_3\left(Q-2p_3(9+\vert\eta_1\vert+\eta_2)\right)+\frac{(p_4^2 +p_5^2)}{6}\right],\\
\rm \kappa&=&\rm 1,\nonumber
\end{eqnarray}
in this case $\sqrt{b}$ is real, then $\rm Z_\rho$ must be the ordinary Bessel function $\rm J_\rho$, therefore
\begin{equation}
\rm {\cal B}_2=Exp\left[
\frac{\left(Q-4p_3(9-\eta_2)\right)}{4\eta_2}\xi_2\right]\,
J_{\rho_2}\left[\frac{2}{\hbar}\sqrt{\frac{2V_{02}}{\eta_2}}
e^{\frac{\xi_2}{2}} \right] \,, \label{b2}
\end{equation}
with  
\begin{equation}\small
\rm \rho_2=\sqrt{\left[\frac{\left(Q-4p_3(9-\eta_2)\right)}{2\eta_2}\right]^2+\frac{1}{\eta_2}\left[\nu^2+p_3\left(Q-2p_3(9+\vert\eta_1\vert+\eta_2)
\right)+\frac{(p_4^2 +p_5^2)}{6}\right]} \,.     
\end{equation}
Finally the wave function $\rm \Psi$ in the original variables
becomes
\begin{equation}\label{psi_caso2}
\rm\Psi=\Psi_0 V^{2\,\alpha} \, Exp\left[p_4 \beta_+ + p_5 \beta_- +
\beta_1\,\lambda_1\phi_1+\beta_2 \lambda_2 \phi_2 \right] \rm
K_{\rho_1}\left[\frac{2}{\hbar}\sqrt{\frac{2V_{01}}{\vert\eta_1\vert}}\, V
e^{-\frac{\lambda_1 \phi_1}{2}}  \right]\,
J_{\rho_2}\left[\frac{2}{\hbar}\sqrt{\frac{2V_{02}}{\eta_2}}\, V
e^{-\frac{\lambda_2 \phi_2}{2}} \right] \,,
\end{equation}
where $\rm \Psi_0$ is a normalization constant, the volume
function $\rm V=ABC=e^{3\Omega}$, and
\begin{eqnarray}
&&\rm \beta_1=\frac{\left(Q-4p_3(9+\vert\eta_1\vert)\right)}{4\vert\eta_1\vert}+p_3 \,,\quad \beta_2=-\frac{\left(Q-4p_3(9-\eta_2)\right)}{4\eta_2} +p_3 \,,\nonumber \\
&&\rm \alpha=-\frac{\left(Q-4p_3(9+\vert\eta_1\vert)\right)}{\vert\eta_1\vert}
+\frac{\left(Q-4p_3(9-\eta_2)\right)}{\eta_2}+4p_3 \,. 
\end{eqnarray}
\begin{figure}[ht!]
\begin{center}
\includegraphics[scale=0.3]{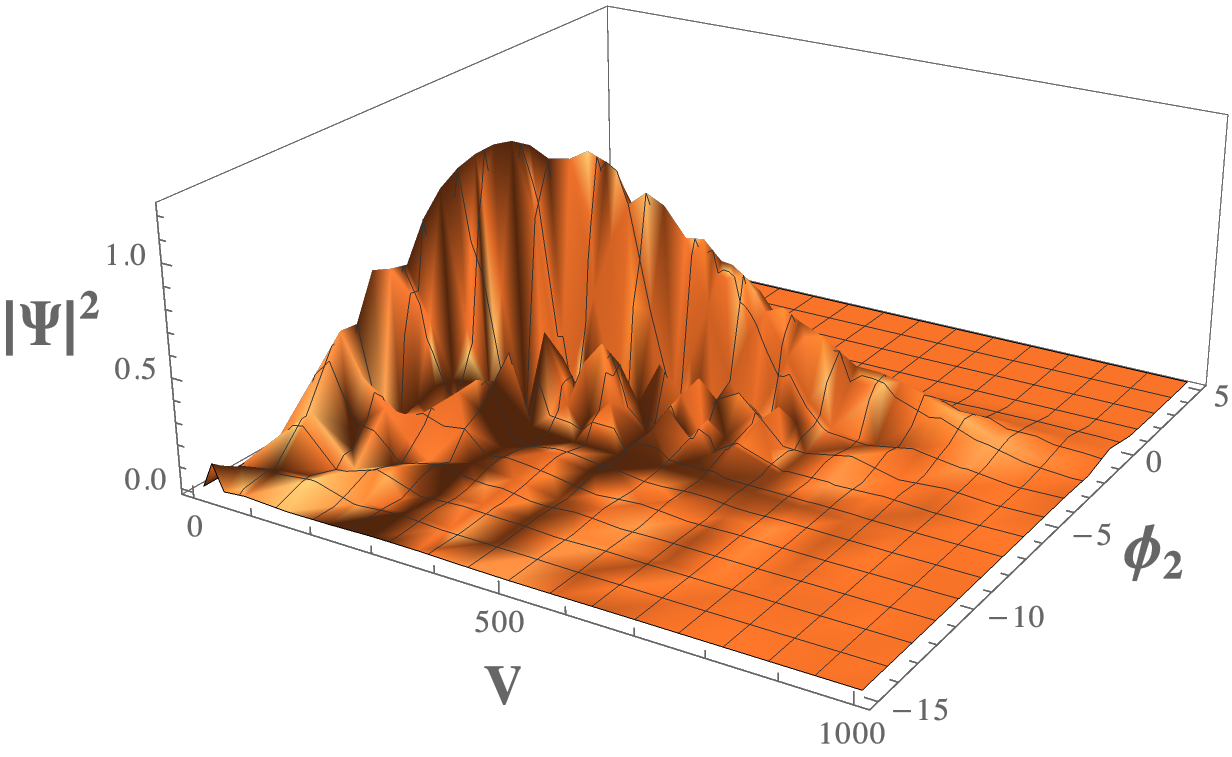}
\includegraphics[scale=0.3]{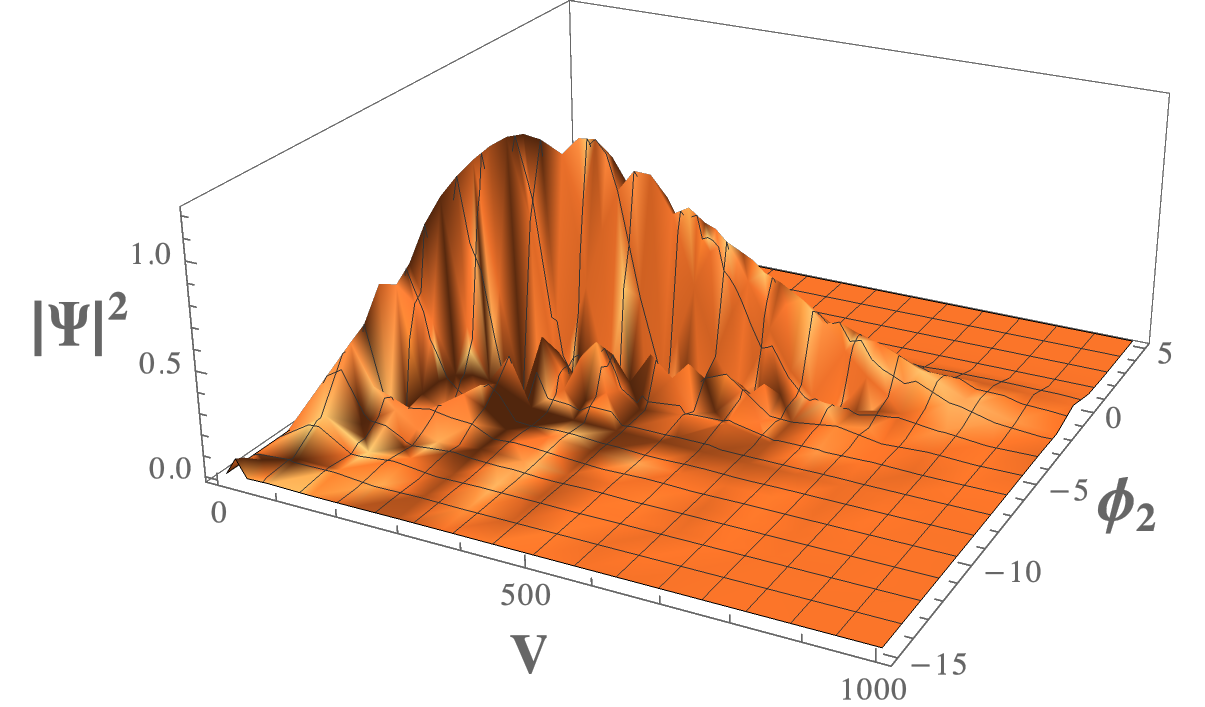}
\includegraphics[scale=0.3]{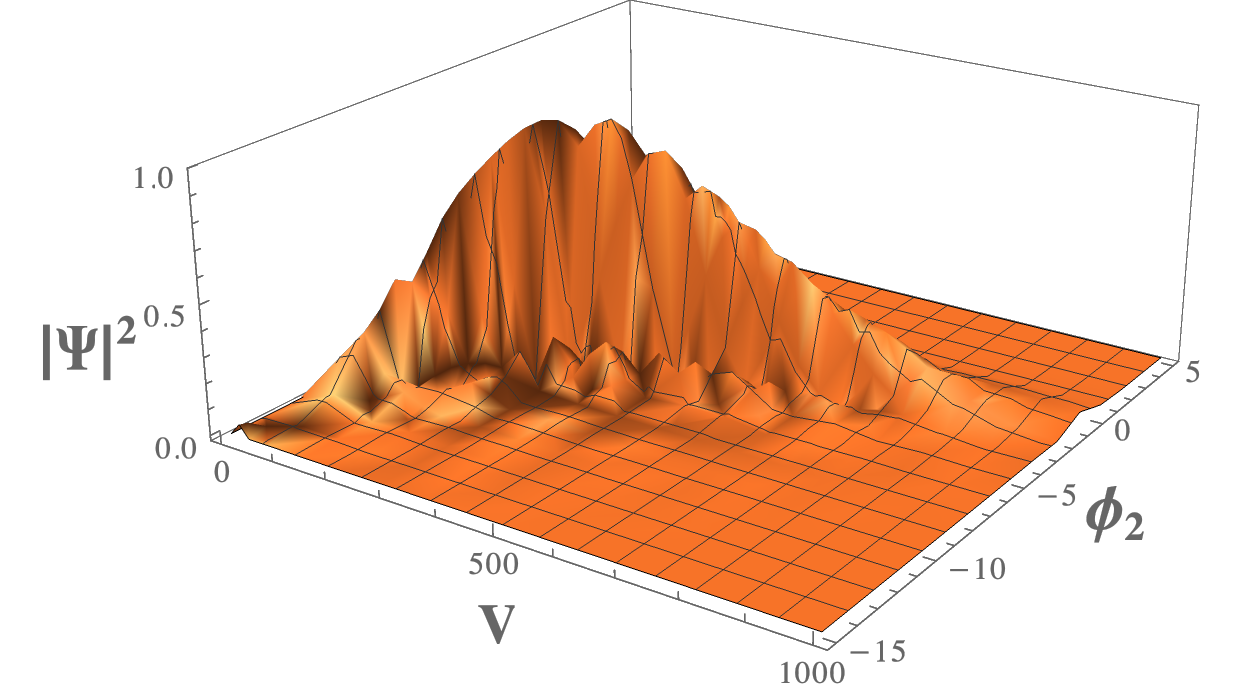}
\includegraphics[scale=0.3]{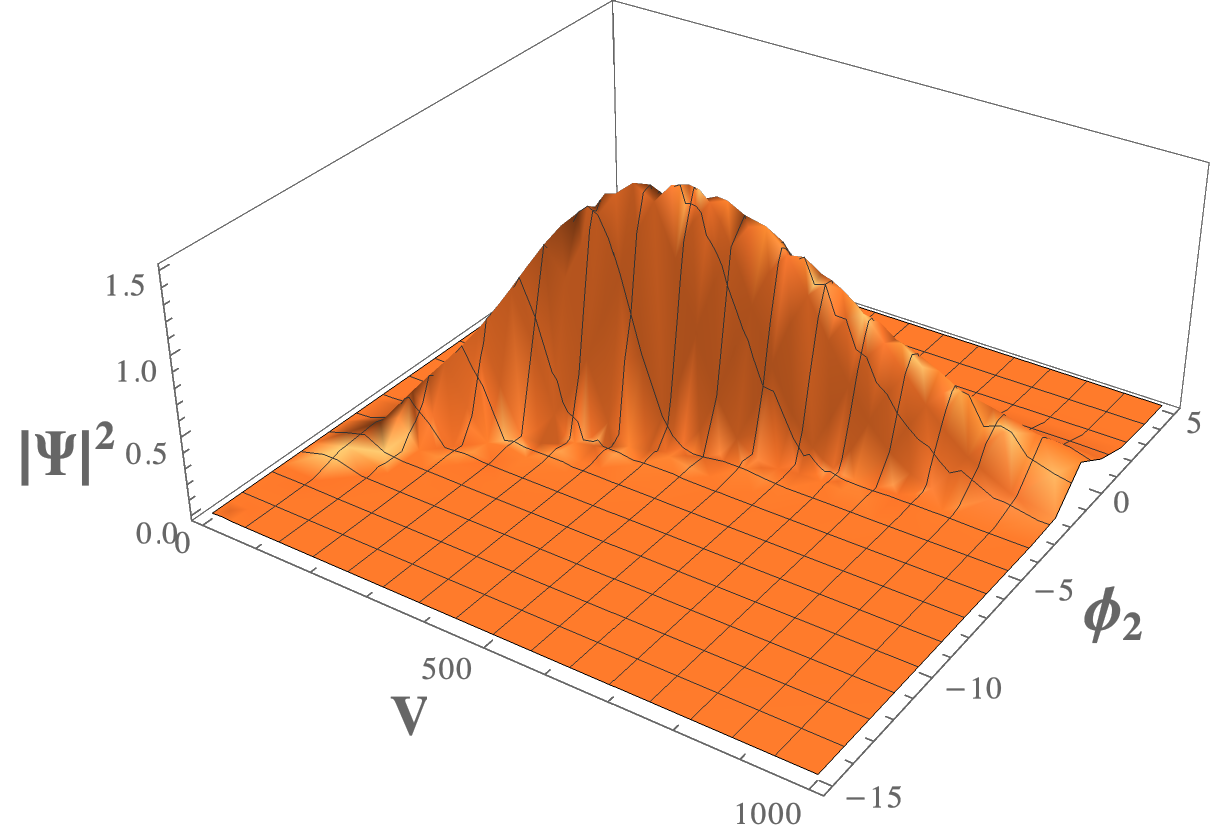}
\caption{Phantom scenario. These figures show the probability
density of the wave function $\rm\vert\Psi\vert^{2}$ (from eq.~(\ref{psi_caso2})) in terms of the volume function $\rm V$ and the scalar field $\rm\phi_{2}$ for various values of $\rm Q=2,0$ (top
panels from left to right), and $\rm Q=-2,-15$ (bottom panels from
left to right). We use arbitrary units of $\nu=10$, $\rm
\lambda_1=4.3\sqrt{6}$, $\rm \lambda_2=6/\lambda_1$, $\rm V_{01} = 5, V_{02} = 10^{-5}$, $\rm r_1=16$, $\rm r_2= 0.5, a_3=-0.5$, $\rm
a_4=a_5=1$, $\phi_1=1$, $\beta_+=\beta_-=1$. Other constants depend on
the aforementioned values.} 
\label{figura_variosQ}
\end{center}
\end{figure}
The behavior of the probability density $\rm\vert\Psi\vert^{2}$ (from eq.~(\ref{psi_caso2})) in terms of the volume function $\rm V$ and the scalar field $\rm\phi_{2}$ is presented in Fig.~\ref{figura_variosQ}. Also, the evolution of $\rm\vert\Psi\vert^{2}$ is shown for different values of the factor ordering parameter $\rm Q$; and actually, in all panels the probability density dies away as $\rm V$ and $\rm\phi_{2}$ evolve. An expected outcome already reported in~\cite{Socorro:2020nsm,Socorro:2022aoz,s-2021,Socorro:2018amv,Socorro:2019wpu}. Moreover, when $\rm Q\ll 0$ then $\rm\vert\Psi\vert^{2}$ tends to behave similar to the isotropic case~\cite{Socorro:2022aoz}.

\subsection{Quantum solution for
$\lambda_1<\sqrt{6}\, \left(\eta_1>0\right)$, and $\lambda_2>\sqrt{6}$.}
For this case equations for $\rm {\cal B}_1$ and $\rm {\cal B}_2$ are similar, therefore they have the same type of solution
\begin{eqnarray}
\rm {\cal B}_1&=& \rm Exp\left[
\frac{\left(Q-4p_3(9-\eta_1)\right)}{4\eta_1}\xi_1\right]\,
J_{\rho_1}\left[\frac{2}{\hbar}\sqrt{\frac{2V_{01}}{\eta_1}}
e^{\frac{\xi_1}{2}} \right] \,, \label{b1-1}\\
 \rm {\cal B}_2&=&\rm Exp\left[
\frac{\left(Q-4p_3(9-\eta_2)\right)}{4\eta_2}\xi_2\right]\,
J_{\rho_2}\left[\frac{2}{\hbar}\sqrt{\frac{2V_{02}}{\eta_2}}
e^{\frac{\xi_2}{2}} \right] \,, \label{b2-2}
\end{eqnarray}
with 
\begin{eqnarray}\small
&&\rm \rho_1=\sqrt{\left[\frac{\left(Q-4p_3(9-\eta_1)\right)}{4\eta_1}\right]^2+\frac{\left[-\nu^2+p_3\left(Q-2p_3(9-\eta_1+\eta_2)
\right)+\frac{(p_4^2 +p_5^2)}{6}\right]}{\eta_1}} \,\\
&&\rm \rho_2=\sqrt{\left[\frac{\left(Q-4p_3(9-\eta_2)\right)}{4\eta_2}\right]^2+\frac{\left[\nu^2+p_3\left(Q-2p_3(9-\eta_1+\eta_2)
\right)+\frac{(p_4^2 +p_5^2)}{6}\right]}{\eta_2}} \,.
\end{eqnarray}
The quantum solution of this set of parameters does not lead to a collapse of the probability density, thus the universe might be eternally quantum and the classical world never takes place. 
\subsection{Quantum solution for
$\lambda_1=\lambda_2=\sqrt{6}\, \left(\eta_1=0\right)$, and $\eta_2=6$.}
Now the equations for $\rm {\cal B}_1$ and $\rm {\cal B}_2$ are reduced to
\begin{equation}
\rm  6\left(Q -36p_3\right)\frac{d {\cal B}_1}{d\xi_1}+3\,\left[-\nu^2+p_3\left(Q-30p_3 \right)+\frac{(p_4^2+p_5^2)}{6}-8\frac{V_{01}}{\hbar^2}
e^{\xi_1}\right]{\cal B}_1=0
\end{equation}
\begin{equation}
\rm -72  \frac{d^2 {\cal B}_2}{d \xi_2^2}+6\left(Q-12p_3\right)\frac{d {\cal B}_2}{d\xi_2}
\rm +3\,\left[\nu^2+p_3\left(Q-30p_3 \right)+\frac{(p_4^2+p_5^2)}{6}-8\frac{V_{02}}{\hbar^2} e^{\xi_2}\right]{\cal B}_2=0 \,,
\end{equation}
and their corresponding solutions are
\begin{eqnarray}
\rm {\cal B}_1 &=& \rm {\cal B}_0
Exp\left[\frac{-\nu^2+p_3(Q-30p_3)+\frac{p_4^2+p_5^2}{6}}{2(36p_3-Q)}\xi_1+\frac{4V_{01}}{\hbar^2(Q-36p_3)}e^{\xi_1}
\right] \,,\\
\rm {\cal B}_2 &=& \rm Exp\left[\frac{Q-12p_3}{24}\xi_2 \right]\,
J_{\rho_2}\left[ \frac{2}{\hbar}\sqrt{\frac{V_{02}}{3}}\,
e^{\frac{\xi_2}{2}}\right] \,,
\end{eqnarray}
with 
\begin{equation}
\rm \rho_2=\sqrt{\left(\frac{Q-12p_3}{12}
\right)^2+\frac{\nu^2+p_3\left(Q-30p_3 \right)+\frac{(p_4^2+p_5^2)}{6}}{6}} \,.    
\end{equation}

\begin{figure}[ht!]
\begin{center}
\includegraphics[scale=0.4]{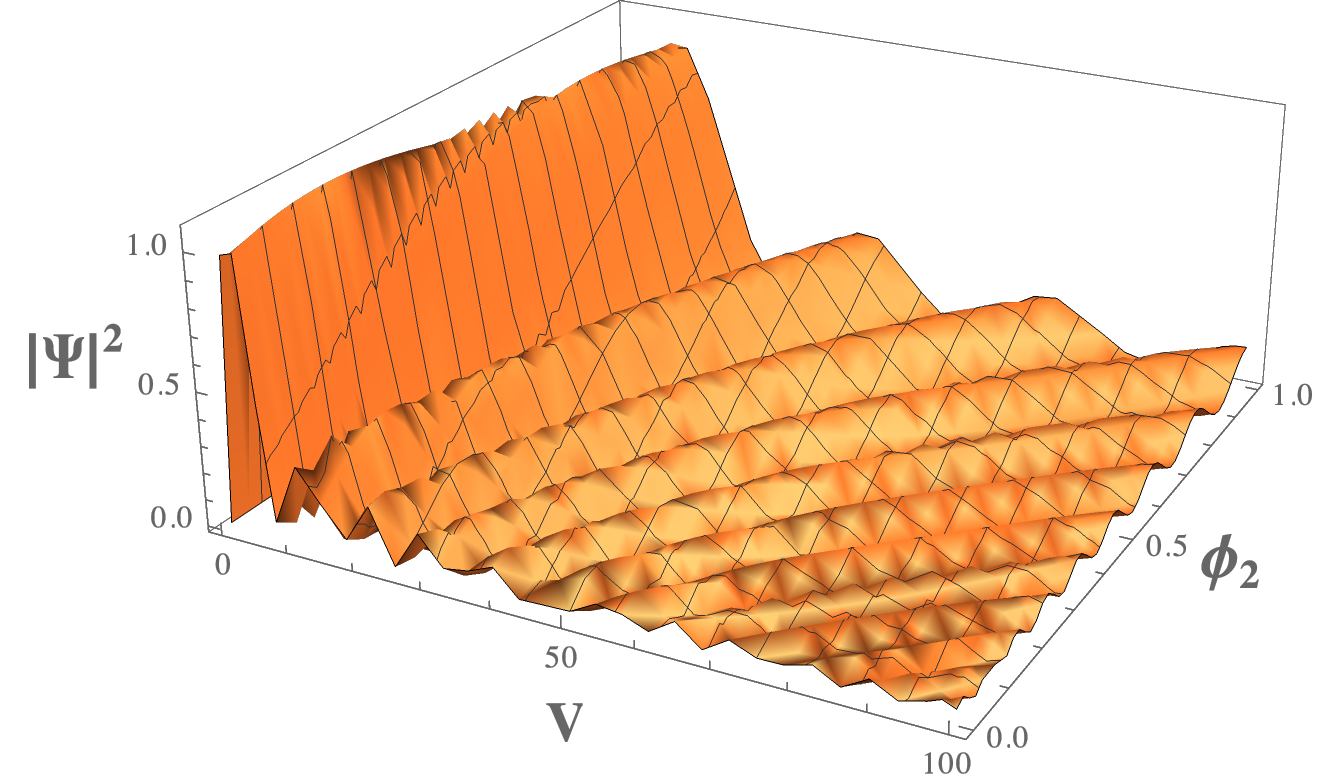}
\caption{This figure shows the probability
density $\rm\vert\Psi\vert^{2}$ of the wave function (\ref{iguales}) in terms of the volume function $\rm V$ and the scalar field $\rm\phi_{2}$. We use arbitrary units, namely $\rm Q=1.5$, $\nu=0.5$, $\rm
\lambda_1=\lambda_2=\sqrt{6}$, $\rm V_{01} = 1$, $\rm V_{02} = 0.1$, $\rm
a_1=a_3=a_4=a_5=1$, $\phi_1=\beta_+=\beta_-=1$ and $\Psi_0=1$.} \label{figura_variosQi}
\end{center}
\end{figure}

Thus, the wave function $\rm \Psi$ in the original variables becomes
\begin{equation}
\rm \Psi=\Psi_0 V^{2\alpha} J_{\rho_2}\left[ \frac{2}{\hbar}\sqrt{\frac{V_{02}}{3}}\, V\,e^{-\frac{\lambda_2}{2}\phi_2}\right]
\rm \times Exp\left[p_4\beta_+ +p_5 \beta_-+\alpha_1 \lambda_1 \phi_1 +\alpha_2 \lambda_2 \phi_2 +\frac{4V_{01}}{\hbar^2(Q-36p_3)}\, V^2\, e^{-\lambda_1 \phi_1}\right] \\
 \label{iguales}
\end{equation}
where $\rm \Psi_0$ is a normalization constant, the volume
function $\rm V=ABC=e^{3\Omega}$, and
\begin{eqnarray}
&&\rm \alpha_1=-\frac{-\nu^2+p_3(Q-30p_3)+\frac{p_4^2+p_5^2}{6}}{2(36p_3-Q)}+p_3 \,, \quad \alpha_2=-\frac{Q-12p_3}{24}+p_3 \nonumber\\
&&\rm \alpha=\frac{-\nu^2+p_3(Q-30p_3)+\frac{p_4^2+p_5^2}{6}}{2(36p_3-Q)}+p_3
+\frac{Q-12p_3}{24} \,.
\end{eqnarray}
Fig.~\ref{figura_variosQi} shows once again a damped $\rm\vert\Psi\vert^{2}$ (from eq.~(\ref{iguales})) due to evolution of $\rm V$ and the scalar field $\rm\phi_{2}$. Now $\rm Q$ compresses the length over the axis on which the scalar field unfolds as time goes by, which in turns delays the progression of the probability density, hence retarding as well the accelerated expansion.
%
\section{Final Remarks}

In this work we have performed a detailed analysis of a chiral anisotropic cosmological model from the perspective of quintom fields. The configuration of our proposal consisted of two fields interact in a non-standard (chiral) way: one quintessence and one phantom; evolving within an anisotropic Bianchi type I background. Both a classical description and its quantum counterpart were presented.

In the classical scenario we find analytical solutions given a particular choice of the emerged relevant parameters. We highlight two cases. The first one when $\rm \lambda_1\lambda_2=\sqrt{6}$, where we have selected the phantom domination instance at $\lambda_1>\sqrt{6}$ ($\eta_1<0$) and $\lambda_2<\sqrt{6}$; and the second one for $\lambda_1=\lambda_2=\sqrt 6$. Then in figures~\ref{figura_1} and~\ref{figura_2} we presented the time evolution of the volume $\rm V$ and the Hubble $\rm H$ functions. Notably in both scenarios the ``big-bang'' singularity is avoided via a ``big-bounce'', yet $\rm V$ grows very rapidly from there. Moreover, the horizontal crossing of the Hubble parameter (at $\rm H=0$) happens at the time of the ``big-bounce'', thus reasserting this result. Indeed, this outcome has been already pointed out in a FLRW framework~\cite{Socorro:2022aoz}. Also, we showed that isotropization is in fact reached as the time evolves for the two examples.   

In the quantum scheme, the WDW equation is constructed and analytically solved for various instances given by the same parameter space from the classical study. First in fig.~\ref{figura_variosQ} we presented four different examples due to $\rm Q$ of the evolution of the probability density $\rm\rm\vert\Psi\vert^{2}$ in terms of $\rm V$ and $\rm\phi_{2}$. All cases display the expected damped behavior as $\rm V$ and $\rm\phi_{2}$ evolve. Remarkably, when $\rm Q \ll 0$ the probability density tends to resemble to the isotropic case~\cite{Socorro:2022aoz}. This result might indicate that for a fixed set of parameters, the anisotropies quantum-mechanically vanish for very small values of the parameter $\rm Q$. Besides, this upshot (at least for small initial anisotropies) was already reported in~\cite{s-2021}. In the second example (fig.~\ref{figura_variosQi}) the probability density also dies away as $\rm V$ and $\rm\phi_{2}$ evolve. However, this time $\rm Q$ compresses the length over the axis on which the scalar field unfolds as time goes by, which in turns delays the progression of the probability density, hence retarding as well the accelerated expansion.

Finally, classical and quantum solutions reduce to their flat FLRW counterparts when the anisotropies vanish.

\begin{acknowledgements}
This work was partially supported by the following grants: J.S. was partially supported by PROMEP UGTO-CA-3 and SNI-CONAHCyT and L. R. D. B. were partially supported SNI-CONAHCyT and Secretaria de Investigación y Posgrado del Instituto Politécnico Nacional, grant SIP20230114. R. H. J. is supported by CONAHCyT Estancias Posdoctorales por M\'exico, Modalidad 1: Estancia Posdoctoral Acad\'emica and SNI-CONAHCyT . S. P. P. was partially supported by  SNI-CONAHCyT and Secretaria de Investigación y Posgrado del Instituto Politécnico Nacional, grant SIP20231773. A. E. G. was partially supported by SNI-CONAHCyT and Secretaria de Investigación y Posgrado del Instituto Politécnico Nacional, grant SIP20231739. This work is part of the collaboration within the Instituto Avanzado de Cosmolog\'{\i}a and Red PROMEP: Gravitation and Mathematical Physics under project {\it Quantum aspects of gravity in cosmological models, phenomenology and geometry of space-time}. Many calculations where done by Symbolic Program REDUCE 3.8.
\end{acknowledgements}
%

\end{document}